# Designing AI-Supported Focus Groups: A Role × Modality Playbook

ZHIQING WANG, University of California, San Diego, USA
STEVEN DOW, University of California, San Diego, USA

Collecting participants' lived experiences is central to design research. Focus groups are uniquely valuable because participants not only share individual accounts but also respond to one another, surfacing comparison, disagreement, and collective sensemaking. However, focus groups are resource-intensive and highly sensitive to facilitation: moderators must probe for specificity, balance participation, manage topic flow, and sustain psychological safety, and subtle facilitation choices can shape what becomes salient. Recent HCI work and commercial meeting tools show that generative AI can scaffold live conversation through prompting, turn regulation, thematic mapping, and real-time summarization. Yet UXR teams lack a clear map of what these capabilities mean in focus groups and what methodological risks they introduce. We synthesize AI supports for live conversation and translate them into a focus-group-specific playbook organized by AI role (tool, co-host, host) and modality (text, voice, embodied). We synthesize prior work on AI-supported live conversation and propose a focus-group-specific playbook of AI supports organized by role (tool, co-host, host) and modality (text, voice, embodied). We characterize interactional trade-offs and identify open questions for evaluating AI-supported focus groups as methodological configurations.

**ACM Reference Format:**
Zhiqing Wang and Steven Dow. 2026. Designing AI-Supported Focus Groups: A Role × Modality Playbook. In *Proceedings of Workshop on Developing an AI-Powered UX Research Point of View (CHI'26 Workshop).* ACM, New York, NY, USA, 9 pages.

### Practitioner Tension: Why Focus Groups Matter, and Why They Break

Collecting stories of participants' lived experiences is central to understanding needs, breakdowns, and opportunities in design. While interviews elicit detailed narratives and reflections that situate experiences in context [11], focus groups add a distinctive advantage: participants not only share their own accounts but also react to and build on the perspectives of others, surfacing points of comparison, disagreement, and shared experience that deepen collective understanding [28]. These group effects make focus groups particularly useful in early-stage discovery, when teams need to understand variance across users and the social dynamics around a product or practice [19].

However, focus groups are resource-intensive because high-quality sessions depend on skilled moderators, and effective facilitation is difficult to do well. Moderators must simultaneously hold two perspectives: on the one hand, they need to understand the designer's intent—what core questions the team is trying to answer and what kinds of stories would be decision-relevant; on the other hand, they must stay attuned to participants' experiences and conversational norms, asking probes that deepen a story without derailing it or signaling judgment [19, 32]. This translation work requires real-time judgment about when to probe, when to let participants respond to one another, and when to move on. Furthermore, facilitation is not neutral. Subtle choices about when to intervene, how to summarize, and which threads to pursue can shape what participants choose to disclose and what the group treats as salient [19, 32]. Group

Authors' Contact Information: Zhiqing Wang, University of California, San Diego, San Diego, USA, zhw131@ucsd.edu; Steven Dow, University of California, San Diego, San Diego, USA, spdow@ucsd.edu.

configuration, especially group size, adds additional constraints and makes facilitation even more delicate. As group size increases, discussion can shift from interactive dialogue to serial monologue, concentrating influence in a few voices and reducing peer-to-peer responsiveness [13]. Related findings in online discussions show that larger groups face coordination costs and social loafing that can lower participation equity and contribution quality [21, 40]. When facilitation is miscalibrated or the group configuration is not well matched to the research goals, the session can lose the key value that distinguishes focus groups from other UX methods: the interactional "group effect," where participants build on, contest, and refine one another's accounts. Instead, the discussion may produce shallow, fragmented, or uneven data—making focus groups expensive to run while yielding limited insight.

### GenAI Supports for Live Conversation Facilitation

Recent HCI work shows that LLM-based systems can scaffold live discussion by intervening in the interactional work that makes conversations productive. Below, We synthesize this literature into recurring mechanisms that are relevant to focus groups and could be adapted to focus-group practice.

**Participation monitoring and equity nudges.** One repeatedly demonstrated capability is real-time monitoring of contribution patterns (e.g., who has spoken, how long, and in what cadence) paired with lightweight nudges that lower participation imbalance. In synchronous group chat, facilitator bots have improved discussion efficiency and encouraged quieter members through prompts and time-structuring [18]. In meetings, systems leverage sensing and behavioral modeling to surface participation signals and reflection cues: post-meeting dashboards can summarize turn-taking and other behavioral measures to increase awareness of inequities, while affect-aware spotlighting can foreground otherwise-missed audience reactions [24, 33]. More interventionist designs embed a co-host agent in the meeting itself: rather than directly seizing the floor, these agents observe, ask permission, and then intervene to support inclusion—a pattern that users may prefer over fully autonomous moderation, even as it introduces new dynamics around authority and deference [15]. Expectation-setting research further shows that *how* the agent is framed (informing vs. explaining vs. offering control) shapes comfort, embarrassment, and acceptance of participation nudges [12].

**Topic/agenda tracking and coverage support.** Another recurring mechanism externalizes conversational state relative to shared goals. In meetings, GenAI-driven "intentionality" probes can synthesize pre-meeting inputs into likely phases, track progress through phases during talk, and adapt interfaces or recommended materials accordingly [30]. At a finer grain, LLM-enabled dialogue mapping systems can surface emerging themes in real time and support revisitation—turning linear talk into navigable structure without requiring a human to manually index the discussion [7]. Complementary interface work shows how language-oriented intent recognition can link conversational speech to actions (e.g., retrieving or sharing relevant artifacts) to reduce coordination friction that otherwise interrupts discussion flow [39].

**Sensemaking scaffolds: structuring, clustering, and gap highlighting.** Beyond tracking, LLMs can help groups make sense of what has been said by organizing content into interpretable intermediate representations. In collaborative ideation, LLM features that suggest connections, cluster contributions, and foreground gaps can broaden exploration and shape what participants discuss next [14, 31]. In meeting contexts, transcript-grounded recap systems produce hierarchical summaries that invite correction and editing [3]. These scaffolds reduce moderators' invisible cognitive load, but they also risk stabilizing interpretation prematurely if AI-produced structure is treated as authoritative.

**Deepening prompts and adaptive follow-ups.** LLMs have also been used to produce context-sensitive probes that increase specificity and narrative detail—a core requirement for focus-group quality. In one-on-one settings, conversational interviewing systems can paraphrase responses (signaling active listening) and generate tailored follow-up questions that pursue ambiguity or request concrete episodes [38]. In interviews, AI assistants can provide real-time suggestions to a human interviewer (or participant) about what to clarify and what thread to pursue next [20]. For focus groups, this points to a design space where probing can be delivered as private prompts or shared questions, with different consequences for peer dynamics and perceived judgment.

**Relational and conflict-aware interventions.** Finally, some systems scaffold the social quality of discussion itself. Relational conversational support can reframe statements, mediate tension, and encourage cooperation across groups, treating communication repair as a first-class design goal [9]. In focus groups, where disagreement can generate insight but also threaten psychological safety, such interventions raise questions about whether AI should preserve productive tension or smooth it.

### Role × Modality as a Design Choice

Prior work demonstrates a growing menu of AI scaffolds for live conversation, but applying these scaffolds to focus groups requires more than selecting features. Focus groups are interaction-sensitive: small changes in who intervenes and how an intervention is delivered can shift disclosure, disagreement, and peer-to-peer uptake. We therefore introduce *Role* × *Modality* as a methodological lens for reasoning about how AI support is positioned in the session (authority) and how it is presented to the group (social presence).

**Role** captures AI's position within the authority hierarchy of the session. As a backstage **Tool**, AI operates without taking the conversational floor, similar to assistant-moderator or note-taker functions such as coverage tracking or probe suggestion. Research on facilitation and qualitative interviewing shows that backstage scaffolding can support moderator cognitive load while preserving participant-driven dialogue [4]. However, even when support remains behind the scenes, it can still shape the trajectory of discussion: suggested follow-up probes and selective paraphrasing can amplify certain narratives while allowing others to fade, and coverage tracking can create subtle pressure to close threads early in order to get through the guide [4, 34]. As a result, tool-level AI may appear non-intrusive on the surface, yet still influence which themes receive depth and which are deprioritized in the resulting data.

As a visible **Co-host**, AI becomes an interactional role whose interventions are part of the shared conversational record. Organizational and group-process research suggests that adding an additional authority figure can shift participation norms, particularly when that figure is perceived as evaluative or expert [23, 26]. In practice, visible monitoring and intervention can change how people take the floor: participants may become more cautious, direct their contributions toward what the authority seems to value, or refrain from disagreeing in order to avoid standing out. Even subtle signals of oversight can reduce dissent and increase conformity pressures [23]. Thus, AI co-hosting can help by regulating turns and making inclusion norms explicit, but it may also increase evaluation apprehension and shape what participants feel safe to share.

As a primary **Host**, AI assumes full facilitation authority—setting the rhythm of turn-taking, deciding when to probe, and establishing what counts as on-topic.This configuration concentrates both interactional control and interpretive power in a single, nonhuman actor, which creates two distinctive risks. First, host authority can amplify errors. When the AI mistimes an interruption, misreads intent, or advances an unhelpful framing, the session may nonetheless follow that trajectory because people often struggle to calibrate reliance on automated partners and can either over-trust or under-trust depending on cues and uncertainty[22, 36, 37]. Work in high-stakes decision support further shows that users do not simply accept or reject AI advice; instead, they may selectively comply, delay, or "negotiate" recommendations—patterns that could interact unpredictably with a host-level facilitator that controls the flow of talk [35]. Second, AI hosting introduces a legitimacy and autonomy challenge. Unlike a backstage tool or co-host, a host configuration can feel like being directed or evaluated by an algorithm, increasing resistance, disengagement, or withdrawal—especially after the system makes visible mistakes. CHI work on algorithm aversion shows that providing people with meaningful forms of control (e.g., outcome or process control) can mitigate aversion [8], suggesting that an AI host without clear override rituals may be particularly brittle. Therefore, while AI hosting may increase standardization and scalability, it also heightens risks of brittle trust dynamics, amplified missteps, and negative affect toward the session leader, potentially reducing the participant-to-participant exchange that focus groups rely on.

**Modality** captures AI's level of social presence *and* the interactional cost of an intervention (i.e., how ignorable vs. interruptive it is). In focus-group facilitation, this matters because the same support (e.g., a recap or a nudge) can function as a quiet backstage cue in one modality, but become an "official" move that reshapes the floor in another. **Text**-based interventions (e.g., side panels, chat prompts) are typically lower-immediacy and easier to ignore or defer, which can preserve conversational flow and reduce perceived evaluation pressure. This aligns with findings from group-chat facilitation work where bots can scaffold participation through lightweight textual prompts without fully taking over the floor [12, 18]. In meeting contexts, the broader CHI/CSCW literature on presence and awareness similarly treats visibility and interactional salience as design variables: systems often struggle to provide social presence without making users feel monitored [17].

**Voice** interventions increase salience, raising the perceived intentionality and authority of the system's moves. As a result, voice-based facilitation can be effective for timing-sensitive coordination (e.g., turn-taking or redirection), but also increases the risk that participants defer to the system's framing or experience the AI as interruptive. CHI work on voice and speech interaction further suggests that vocal qualities shape perceived credibility and listening experience, implying that voice delivery is not a neutral channel for facilitation [5].

**Embodied** or avatar-based agents further elevate social presence by adding a visible "someone" into the interaction, increasing norm awareness and self-monitoring. In mixed-modality conferencing, introducing avatars changes co-presence and perceived presence of others, and highlights how agency/visual realism modulate social experience [29]. Similarly, CHI work on embodied conversational agents shows that adding expressive, style-matching embodiment can increase perceived empathy and believability, reinforcing that embodiment alters how interventions are interpreted [2]. For focus groups, these results suggest a practical tradeoff: higher-presence modalities can strengthen engagement and coordination, but they also heighten risks of deference, impression management, and disruption when facilitation moves are mistimed or perceived as evaluative.

By crossing Role and Modality, we identify distinct methodological profiles for AI-supported focus groups. Each configuration introduces predictable benefits and risks grounded in contemporary research on qualitative facilitation, group dynamics, trust calibration, and social presence.

## Playbook: What AI Can Do in a Live Focus Group

Building on the Role × Modality framework, Table 2 translates common GenAI conversational capabilities into focus-group-specific facilitation plays. Rather than treating AI support as a monolithic feature set, we frame each capability as an interactional intervention that carries predictable effects on participation, narrative depth, coordination, and psychological safety.

Each play specifies (1) the implied Role and Modality configuration, (2) the mechanism by which it reshapes group interaction, and (3) risks to interaction validity such as anchoring, deference, interruption cost, spotlighting, and premature convergence. Importantly, these plays can be implemented at different authority and presence levels; the same capability may function as low-risk backstage scaffolding (Tool × Text) or as high-impact authoritative intervention (Host × Voice).

Table 1. Role × Modality configurations and their likely effects on group interaction.

| Role | Text | Voice | Embodied |
|---|---|---|---|
| **Tool** | **Support:** Guides the moderator privately (probe suggestions, topic coverage checks, recap drafts) without taking speaking turns.<br>**Risk:** Still shapes what gets explored: suggested probes and selective recaps can pull attention toward some themes and away from others.<br>[3, 4, 34, 38] | **Support:** Provides backstage timing support (live transcription, pacing cues) that helps keep the session on track.<br>**Risk:** Makes monitoring more noticeable; participants may feel "watched" and become less spontaneous.<br>[6, 33] | **Support:** Adds subtle awareness signals (e.g., status indicators) that help the moderator track progress.<br>**Risk:** A visible "system presence" can increase surveillance concerns even if the AI does not speak.<br>[17] |
| **Co-host** | **Support:** Posts visible nudges and summaries that encourage participation and keep discussion organized, while humans still control the floor.<br>**Risk:** Summaries can become an "official" framing that anchors what the group treats as important; nudges can also spotlight individuals.<br>[7, 12, 18] | **Support:** Speaks to regulate turns and rebalance participation in the moment (e.g., redirecting, inviting quieter members).<br>**Risk:** Spoken interventions carry more authority and interruption cost; participants may defer to the AI's framing or hold back disagreement.<br>[15, 22] | **Support:** A visible co-facilitator can strengthen coordination by making norms and feedback cues more explicit.<br>**Risk:** Higher social presence increases self-consciousness; participants may manage impressions instead of speaking freely.<br>[2, 24, 29] |
| **Host** | **Support:** Runs a consistent discussion structure at scale (standardized sequence, probes, and transitions).<br>**Risk:** Errors in framing or timing have larger impact because the AI controls the agenda; this can reduce peer-to-peer exchange.<br>[1, 8, 37] | **Support:** Directly controls turn allocation and pacing, which can reduce chaos and keep the group moving.<br>**Risk:** Trust can drop quickly after visible mistakes; participants may feel less agency and speak less, especially if they cannot easily override the AI.<br>[10, 26, 36] | **Support:** High presence can boost engagement and make facilitation cues hard to miss.<br>**Risk:** Strong authority signals can increase norm pressure and evaluation concerns, leading to guarded disclosure or withdrawal.<br>[25, 27] |

## Future Work

The integration of generative AI into focus groups is not merely a tooling question; it is a methodological one. Rather than asking whether AI can facilitate conversation, future research should investigate how specific Role × Modality configurations systematically reshape the qualitative data that designers ultimately interpret.

First, controlled empirical comparisons are needed across configurations (e.g., Tool vs. Co-host vs. Host; Text vs. Voice). Prior work on conversational cueing and constructive communication suggests that subtle scaffolds can shift participation balance and topical diversity [16, 31]. Extending this logic to focus groups, future studies should examine how AI support influences established qualitative dimensions such as topic breadth, narrative specificity, interactional reciprocity, and designer-rated usefulness. Importantly, comparisons should evaluate not only output quantity but interactional dynamics: who speaks, who defers, and how disagreement unfolds.

Second, authority calibration requires systematic investigation. Our framework suggests that as AI moves from Tool to Host, authority and interruption cost increase. Research in trust calibration and AI-assisted decision-making shows that visibility and control shape deference and reliance [8, 22]. In focus groups, similar dynamics may alter disclosure patterns, dissent, and willingness to challenge emerging interpretations. Future work should explore how transparency mechanisms, opt-out controls, or explicit framing strategies mitigate authority amplification while preserving coordination benefits.

Table 2. AI Support Playbook for Focus Groups

| Play Card | Interactional Mechanism | Configuration, Risks, Related Work |
|---|---|---|
| Deepening Story Prompts | Suggests context-sensitive follow-ups that encourage elaboration of specific moments, increasing narrative specificity and contextual richness. | **Role:** Tool or Co-host<br>**Modality:** Text (low interruption) or Voice (higher salience)<br>**Risks:** Framing bias; narrowing narrative scope; anchoring interpretations<br>**Related Work:** Adaptive interviewing [38]; Interview AI-ssistant [20]. |
| Equitable Turn Nudges | Detects participation imbalance and invites quieter members to contribute, redistributing floor time. | **Role:** Co-host<br>**Modality:** Text (gentle nudge) or Voice (explicit invitation)<br>**Risks:** Spotlight pressure; perceived monitoring; self-censorship<br>**Related Work:** Bot in the Bunch [18]; AI meeting co-host [15]. |
| Topic Coverage Monitoring | Tracks agenda alignment and flags underexplored themes, shaping pacing and closure decisions. | **Role:** Tool<br>**Modality:** Text (side-panel indicators)<br>**Risks:** Over-structuring; premature topic shifts; reduced emergence<br>**Related Work:** MeetMap [7]; Conversational cueing [31]. |
| Live Thematic Reflection | Generates interim summaries and prompts validation or correction, externalizing group interpretation in real time. | **Role:** Tool or Co-host<br>**Modality:** Text (shared recap)<br>**Risks:** Anchoring; premature consensus; authority amplification if treated as official<br>**Related Work:** Dialogue mapping [7]; LLM recap systems [3]. |
| Relational Reframing | Reinterprets disagreement as contrasting needs or perspectives to sustain constructive exchange. | **Role:** Co-host or Host<br>**Modality:** Text or Voice<br>**Risks:** Sanitizing productive conflict; privileging system framing<br>**Related Work:** Relational AI [9]. |
| Idea Clustering and Gap Highlighting | Groups emerging stories into thematic clusters and identifies missing viewpoints, shaping collective sensemaking. | **Role:** Tool<br>**Modality:** Text (visual canvas / summary layer)<br>**Risks:** Premature convergence; reduced interpretive diversity<br>**Related Work:** AI ideation canvas [14]; Cueing the Crowd [31]. |
| Structured Turn Management | Allocates turns or enforces structured rounds to reduce dominance and maintain coordination. | **Role:** Host (high authority)<br>**Modality:** Voice (high interruption cost)<br>**Risks:** Deference; reduced spontaneity; amplified errors if mistimed<br>**Related Work:** AI meeting facilitation agents [15]; LLM-facilitated group decision systems [1]. |

Finally, evaluation frameworks tailored to AI-supported qualitative research must be developed. Beyond efficiency metrics, we need measures of interaction validity: Does AI increase story concreteness? Does it broaden or narrow interpretive diversity? Does it sustain the interactional “group effect” that distinguishes focus groups from serial interviews? Developing such criteria would allow AI-supported focus groups to be assessed as methodological innovations rather than productivity enhancements.

In sum, future work should treat AI support as part of the method: different Role × Modality choices are likely to produce different participation patterns and different types of stories. These directions position AI-supported focus groups not as automation of moderation, but as a new design space for methodological configuration—where authority, presence, and scaffolding are variables to be intentionally calibrated rather than implicitly inherited from human facilitation.

## Conclusion

Focus groups are uniquely valuable because they generate insight through interaction. Participants do not merely report experiences; they respond to, challenge, and build on one another’s stories. Yet this richness depends heavily on facilitation, making focus groups both powerful and fragile as a research method. Generative AI is now applied in live conversational settings and taking on functions that are part of facilitation, from prompting participation to managing topic flow and summarizing emerging themes. These capabilities present new opportunities for supporting qualitative research, but they also introduce methodological risks related to authority, framing, convergence, and disclosure. In this paper, we synthesized existing research on AI-supported conversation and translated it into a focus-group-specific playbook organized by role and modality. By framing AI integration as an interaction design decision with epistemic consequences, we aim to provide UXR practitioners with a structured way to reason about when and how to incorporate AI into focus group practice. AI will not replace the need for methodological care. Instead, it creates a new layer of design choices that shape how stories are elicited, interpreted, and transformed into insight. Understanding these choices is essential if AI-supported focus groups are to advance, rather than undermine, qualitative design research.